\documentstyle[prl,aps]{revtex}
\newcommand{\BEQ}{\begin{equation}}
\newcommand{\EEQ}{\end{equation}}
\newcommand{\BEA}{\begin{eqnarray}}
\newcommand{\EEA}{\end{eqnarray}}
\begin{document}
\draft
\title{A puzzle on fluctuations of weights in spin glasses}
 \author{Th.~M.~Nieuwenhuizen}
 \address{
 Van der Waals-Zeeman Institute, Universiteit van Amsterdam\\
	 Valckenierstraat 614/5, 1018 XE Amsterdam, The Netherlands\\
e-mail: nieuwenh@phys.uva.nl\\}
\date{12-4-95; revised 29-9-95; submitted to Journal de Physique I}
\maketitle
\begin{abstract}
In certain mean field models for spin glasses
there occurs a one step replica symmetry breaking pattern.
As an example of general $1/N$-corrections in such systems,
the fluctuations in the internal energy are calculated.
For this specific quantity the outcome is known
from the specific heat.
It is shown that the correct result can be derived by
assuming that the both the height and the location of the breakpoint
fluctuate. This effect enlarges the commonly considered
space of allowed fluctuations of the order parameter
in loop calculations for short range spin glasses of this type.
The phenomenon does not occur in spin glasses with infinite
order replica symmetry breaking.
\end{abstract}
\pacs{7510.Nr,7540.Cx, 7550.Lk,8710}

\section{Introduction}
The theoretical understanding of the spin glass phase relies on the
Parisi replica symmetry breaking scheme.\cite{Parisi}
Parisi  considered a variational
problem with $k$ breakings of replica symmetry. The objects that show up are
the values of possible overlaps $q_i$ ($0\le i\le k$)
and their cumulative weights $x_i$  $(1\le i\le k$).
The free energy was maximized in these variables. In the
Sherrington-Kirkpatrick model
the physical result arises in the limit $k\to\infty$. Eliminating
the $i$'s a continuous order parameter function $q(x)$ arises.
Loop corrections to this
mean field theory have been studied in great detail in a series of
papers by de Dominicis, Kondor and Temesvari.~\cite{dDKT}
In particular, the
fluctuation matrix has been inverted in full glory.\cite{dDKTgf}

Many aspects of this and related replica analyses are ill defined and
often ill understood.
Nevertheless replica results for the SK-model have been rederived
by TAP-analysis and cavity methods.~\cite{MPV}
For the random energy model (``simplest spin glass'') the
 replica results are in full agreement with the direct evaluation of
the partition sum.~\cite{GM}

It is our belief that a replica analysis
can produce the physically correct answer in a quick way,
but its interpretation should always
be understood from a more profound analysis. If the anwer is
unsatisfactory, then one should simply abandon it or improve the
derivation (the celeber case being Parisi's improvement of
the Sherrington-Kirkpatrick saddlepoint.)
Such a situation occurs in particular in systems which exhibit one
step of replica symmetry breaking (1RSB). Here there occur two transition
temperatures, $T_g$ for the statics and $T_c>T_g$ for the dynamics.

Kirkpatrick and Wolynes~\cite{KirkpW} considered the Potts glass,
which exhibits a 1RSB when the number of components exceeds four.
In their analysis of the metastable states
(TAP states) they observed that there occurs a critical
temperature $T_c>T_g$ where unexpected behavior sets in.
On a static level the system condenses for $T_g<T<T_c$ in a
multitude of states with
extensive complexity, all of which have non-zero self-overlap but
vanishing mutual overlap. The free energy is therefore still equal to the
paramagnetic free energy. As a result, this transition is not noticed
in a direct replica analysis of the partition sum when one uses
Parisi's prescriptions. Nevertheless, it can be described by
replica's if one modifies the rule for fixing the breakpoint.
The value of the selfoverlap can be obtained by solving the equation
$\partial F/\partial q_1=0$,
with the additional rule that the limit $x_1\to 1$ is taken afterwards.
In this temperature interval, namely, it
gives the maximal free energy in the physical range $0\le x_1\le 1$.
Loosely speaking, that limit equates the value of the selfoverlap
to the mutual overlaps in a limiting small cluster of states.
 This procedure also automatically implies that the free
energy of this phase is equal to the paramagnetic one.
For $T<T_g$ the extremum can no longer be located at $x_1=1$
and the additional restriction has to be relaxed.

A related mystery of the replica calculus occurs in the dynamical
approach to the systems with 1RSB. The prototype is the
spherical $p$-spin interaction spin glass, see eq. (1) below.
The long-time limit of dynamics has a vanishing fluctuation eigenvalue,
and is therefore called ``marginal''.~\cite{KirkpT}~\cite{CHS}~\cite{CK}.
 The long-time dynamical values for $q_1$ and $x_1$
can be derived from the replica free energy by assuming marginality
within the replica calculation~\cite{KirkpT}~\cite{CHS}~\cite{Nmaxmin}.
The mystery remains why a replica calculation can explain
the long-time dynamics. After all, the Gibbs weight is thought to be
related to exponentially large times where the system has time enough
to visit all states. Marginality is thought to be related to
algebraically large times. It therefore seems paradoxical that Gibbs weight at
marginality does describe the results of long time dynamics.
It has been suggested that an essential role is played here by the
complexity. It is so large that the replica free energy is the
lowest of the available states, an old thermodynamic law,
but new in glassy systems.~\cite{Ncompl}

In this note we wish to draw attention to another aspect of the location
of the break points of 1RSB solutions. This time we consider the more
standard stationary case, where the Parisi description applies,
namely the optimal value of the break points follows by maximizing
the free energy.
When calculating $1/N$ contributions to a physical quantity,
we find that care should be taken
of fluctuations in the location of the plateaus.

In next section we present in detail the calculation of the
fluctations of the  internal energy in a $p$-spin interaction
mean field spin glass model with 1RSB. Subsequently this analysis is
generalized to models with $\infty$RSB. The paper closes
with a discussion.

\section{Analysis of a  model with 1RSB}

 For a system with $N$ spins in zero field
we consider the Hamiltonian
\begin{equation}\label{Ham=}
{\cal H}=-\sum_{i_1<i_2<\cdots<i_p} J_{i_1 i_2\cdots i_p}
S_{i_1}S_{i_2}\cdots S_{i_p}
\end{equation}
with independent Gaussian random couplings, that have average zero
and variance $J^2p!/2N^{p-1}$. The spins are spherical and subject to the
condition $\sum_i S_i^2=N$. A replica calulation was performed by
Crisanti and Sommers (CS).~\cite{CS} The replicated free energy reads
\begin{equation}
 F_n=-\frac{\beta J^2}{4}\sum_{\alpha\beta}q_{\alpha\beta}^p
-\frac{T}{2}{\rm tr}_{\alpha}\log q
\end{equation}

 At zero field the order parameter
function has the one-step form $q(x)=q_1\theta(x-x_1)$,
and the  free energy reads
\begin{eqnarray}\label{bFCS}
\frac{ F}{N}=-\frac{\beta J^2}{4}+\frac{\beta J^2}{4}
\xi_1 q_1^p
-\frac{T}{2x_1}\log(1-\xi_1 q_1)+\frac{T\xi_1}{2x_1}\log(1-q_1)
\end{eqnarray}
where $\xi_1=1-x_1$.
 The internal energy reads
\BEQ U=-\frac{\beta J^2N}{2}(1-\xi_1 q_1^p) \EEQ
Following Crisanti and Sommers in their TAP-approach~\cite{CSTAP},
we denote by $z$ the value $0<z<\sqrt{2/p(p-1)}$
where the complexity
\BEQ {\cal I}=\frac{N}{2}\left(\frac{2-p}{p}-\log\frac{pz^2}{2}
+\frac{p-1}{2}z^2-\frac{2}{p^2z^2}\right)\EEQ
 is non-extensive, viz. ${\cal I}/N=0$. The equations
 $\partial F/\partial q_1=\partial F/\partial x_1=0$
then can be cast in the form
\BEQ \beta (1-q)q_1^{p/2-1}=z; \qquad x_1=
\frac{(2-pz^2)}{pz^2}\, \frac{1-q_1}{q_1}
 \EEQ
{}From these results one can calculate the specific heat $C=dU/dT$ using
\BEQ \frac{dq_1}{dT}=\frac{2\beta q_1(1-q_1)}{p-2-pq_1}\EEQ
One obtains
\BEQ \label{C=} C=\frac{\beta^2J^2N}{2}
\left(1-\xi_1 q_1^p-2q_1^p\frac{1-\xi_1(pq_1-p+1)}{pq_1-p+2}\right)\EEQ

Let us now consider the fluctuations in the internal energy,
\BEQ
\Delta U^2\equiv \langle {\cal H}^2\rangle-\langle {\cal H}\rangle^2=
\frac{1}{n} \sum_{\alpha,\beta=1}^n\int DS
{\cal H}(S_\alpha){\cal H}(S_\beta)\exp\left(-\beta\sum_{\gamma=1}^n{\cal H}
(S_\gamma)\right)\EEQ
where the last relation involves replicas, and
the limit $n\to 0$ is implied.
The leading terms in $N$
are of order $N^2n$, so they donot survive for $n\to 0$.
Indeed, $U$ is expected to  have fluctuations of relative
order  $1/\sqrt{N}$.

The first ${\cal O}(N)$ contribution
to $\Delta U^2$ comes when all sites
in the pre-exponentials overlap pairwise. It yields
\BEQ \label{U1}
\Delta U^{2\,(1)}=\frac{J^2N}{2n}\sum_{\alpha\beta}q_{\alpha\beta}^p=
\frac{J^2N}{2}(1-\xi_1q_1^p)
\EEQ
Before calculating the other terms one has to average over the couplings.
It implies the shift
\BEQ J_{i_1\cdots i_p}\to  J_{i_1\cdots i_p}+\frac{\beta J^2 p!}{2N^{p-1}}
\sum_\gamma S_{i_1}^\gamma\cdots S_{i_p}^\gamma\EEQ
One gets the contribution
\BEQ \label{lange}
\Delta U^{2\,(2)}
=\frac{\beta^2J^4N^2p!p!}{4N^{2p}n}
\sum_{\alpha\beta\gamma\delta}
\sum_{i_1<i_2<\cdots<i_p}
\sum_{i_1'<i_2'<\cdots<i_p'}\int \frac{dpdq}{2\pi i}
\int DS
S_{i_1}^\alpha S_{i_1}^\gamma\cdots S_{i_p}^\alpha S_{i_p}^\gamma
S_{i_1'}^\beta S_{i_1'}^\delta\cdots S_{i_p'}^\beta S_{i_p'}^\delta
e^{-\beta {\cal H}_n} \EEQ
where the replicated exponential reads
\BEQ e^{-\beta {\cal H}_n}=\exp\sum_{\alpha\beta}\left[
\frac{N}{4}\beta^2J^2 q^p_{\alpha\beta}+\frac{N}{2}p_{\alpha\beta}
q_{\alpha\beta}-\frac{1}{2}
\sum_i p_{\alpha\beta}S_i^\alpha S_i^\beta\right]\EEQ
and $dpdq/2\pi i$ is a short hand for the integration measure
$\prod_{\alpha<\beta}[dp_{\alpha\beta}dq_{\alpha\beta}/2\pi i]
\prod_\alpha [dp_{\alpha\alpha}/2\pi i]\times (1+{\cal O}(n))$, while further
$p$ and $q$ are symmetric.\cite{CS}
It is indeed seen that the leading terms in $N$ are of order $N^2n$,
so they donot contribute when $n=0$.

We can now express the prefactors as $q$'s and carry out the spin integrals,
to obtain
\BEQ \label{langer}
\Delta U^{2\,(2)}=\frac{\beta^2J^4N^2}{4n}
\int \frac{dpdq}{2\pi i}
\sum_{\alpha\beta\gamma\delta}
\left[q_{\alpha\gamma}q_{\beta\delta}\right]^p
 \exp\left[\sum_{\alpha\beta}\left\{
\frac{N}{4}\beta^2J^2 q^p_{\alpha\beta}+\frac{N}{2}p_{\alpha\beta}
q_{\alpha\beta}\right\}-\frac{N}{2}{\rm tr}\log{p}\right]\EEQ
Expanding around the saddlepoint one sets
$q_{\alpha\beta}=\bar q_{\alpha\beta}+\delta q_{\alpha\beta}$,
$p_{\alpha\beta}=\bar p_{\alpha\beta}+\delta p_{\alpha\beta}$,
with $\delta q_{\alpha\alpha}=0$, where $\bar q$ and
 $\bar p=\bar q^{-1}$ momentarily denote the
saddle point values. One can then integrate out the $\delta p$'s, so that
only the fluctuations in the $q$'s remain
 \BEQ \label{DU22} \Delta U^{2\,(2)}=\frac{\beta^2 J^4 N^2}
{4n [{\rm det}(\bar p\bar p)]^{1/2}}\int dq
\sum_{\alpha\beta\gamma\delta}
q_{\alpha\gamma}^p
q_{\beta\delta}^p\exp\frac{-N}{2}
\sum_{\alpha\beta\gamma\delta}
\delta q_{\alpha\beta} M_{\alpha\beta;\gamma\delta}
\delta q_{\gamma\delta}\EEQ
where $M_{\alpha\beta;\gamma\delta}=\partial ^2\beta F_n/\partial
q_{\alpha\beta}\partial q_{\gamma\delta}$ is the fluctuation matrix.
Its matrix elements and eigenvalues are discussed
by Crisanti and Sommers~\cite{CS}.

Now one  can go to eigenmodes of $M$.~\cite{CS}
The transversal ones can be integrated
out, which brings the square root of a another determinant as prefactor.
At fixed break point $x_1$ only the longitudinal eigenvector
$\delta q_{\alpha\beta}\sim q_{\alpha\beta}$ will contribute due to the
contraction in the preexponential. However, also $q$'s with a
small shift in the breakpoint will be seen to contribute.\cite{Ferrero}
We therefore consider fluctuations described by order parameter functions
\BEQ \label{qform}
q_{\alpha\beta}\to q(x)=(1+\eta)q_1\theta(x-x_1-\delta)\EEQ
where $q_1$ and $x_1$ are the saddlepoint values, and
where $\eta$ gives fluctuations in the height of the plateau and
$\delta$ in the location of the breakpoint.

The shape (\ref{qform}) of $q(x)$ leads to $n$-times the free energy
(\ref{bFCS}) with $q_1\to q_1(1+\eta)$ and $x_1\to x_1+\delta$.
The value at $\delta=\eta=0$ is of order $n$ and has been omitted
already from eq. (\ref{DU22}).
The linear terms vanish due to the saddlepoint condition, and were
also omitted already. The quadratic terms therefore yield immediately
\begin{equation}
\sum_{\alpha\beta\gamma\delta}
\delta q_{\alpha\beta} M_{\alpha\beta;\gamma\delta}
\delta q_{\gamma\delta}\to
n\left(\eta^2q_1^2\beta F_{qq}+2\eta\delta q_1\beta F_{qx}
+\delta^2\beta F_{xx}\right)
\EEQ

We now  assume that for $n\to 0$ the summation
over different breakpoints can be replaced by an integral over their
location $x_1+\delta$ viz. $\int w_1(x_1+\delta)d(x_1+\delta)$, where
$w_1$ is an unknown but smooth weight.
Since $\delta$ is of order $1/\sqrt{N}$,
it then follows that its weight $w_1(x_1+\delta)\approx w(x_1)$
is uniform in $\delta$. The total weight $w$ consists of $w_1$
and of square roots of determinants that arise after performing the
$\delta p$ integrals and the transversal $\delta q$ modes.
Next we assume that $w$ is such that the remaining Gaussian integral over
$\delta$ and $\eta$ is normalized to unity.
One then has
 \BEQ \Delta U^{2\,(2)}=\frac{\beta^2 J^4 N^2n}{4}\,\,
\frac {\int d\eta d\delta
[1-\xi_1 q_1^p+\delta q_1^p-p\xi_1\eta q_1^p]^2
\exp\frac{-Nn}{2}\left(\eta^2q_1^2\beta F_{qq}+2\eta\delta q_1\beta F_{qx}
+\delta^2\beta F_{xx}\right)}
{\int d\eta d\delta
\exp\frac{-Nn}{2}\left(\eta^2q_1^2\beta F_{qq}+2\eta\delta q_1\beta F_{qx}
+\delta^2\beta F_{xx}\right)}
\EEQ
Let us recall that the appearence of the integral in the denominator
is due to our assumed value of $w$, now seen to be of order $Nn$.
As the Gaussian fluctuations bring a factor $1/Nn$, they indeed
yield terms that survive in the limit $n\to 0$.
There are several fluctuation contributions. We should only
take into account the cross terms $[\delta q_1^p-p\xi_1\eta q_1^p]^2$.
This yields
\BEQ \label {19} \Delta U^{2\,(2)}=\frac{\beta^2 J^4N q_1^{2p-2}}{4}
\left(p^2\xi_1^2R_{qq}-2p\xi_1q_1R_{qx}+q_1^2R_{xx}\right)
\end{equation}
where $R$ is the inverse fluctuation matrix,
viz. $R^{-1}_{ij}=\beta F_{ij}$, with $i=q,x$; $j=q,x$.
Inserting the  field equations $\partial F/\partial q_1=
\partial F/\partial x_1=0$ one has
\BEA \beta F_{qq}&=&
\frac{-\xi_1[(p-1)q_1-p+2 +\xi_1 q_1(p-1-pq_1)]}
{2(1-q_1)^2(1-\xi_1 q_1)^2}\\
\beta F_{qx}&=&\frac{\xi_1 q_1^2}{2(1-q_1)(1-\xi_1 q_1)^2}\\
\beta F_{xx}&=&\frac{-q_1^2(pq_1-2\xi_1 q_1-p+2)}
{2p(1-q_1)x_1(1-\xi_1 q_1)^2}\EEA
The inverse matrix reads
\BEA R_{ij}&=&\frac{-2(1-q_1)(1-\xi_1 q_1)}{(2-p+pq_1)
[(p-1)q_1-\xi_1 q_1-p+2]}S_{ij}\\
S_{qq}&=&\frac{1-q_1}{\xi_1}(pq_1-2\xi_1 q_1-p+2)\\
S_{qx}&=&px_1(1-q_1)\\
S_{xx}&=&\frac{px_1}{q_1^2}[(p-1)q_1-p+2 +\xi_1 q_1(p-1-pq_1)]\EEA
Now one only has to insert these results in (\ref{19}). One obtains
\BEA \Delta U^{2\,(2)}
=-\frac{J^2Nq_1^p[1-\xi_1(pq_1-p+1)]}{pq_1-p+2}\EEA
Note that several unpleasant denominators have canceled, affirming that
that our procedure is correct. Putting things together we find
\BEA
\Delta U^2
=\Delta U^{2\,(1)}+\Delta U^{2\,(2)}
=\frac{J^2N}{2}\left(1-\xi_1 q_1^p-2q_1^p
\frac{1-\xi_1(pq_1-p+1)}{pq_1-p+2}\right)
\EEA
Comparing with eq. (\ref{C=}) one sees that indeed $\Delta U^2=T^2dU/dT$,
as required by a standard thermodynamic relation. If variations in
$x_1$ had not been taken into account,  $\Delta U^2$ would have been
proportinal to $1/F_{qq}$, which is certainly incorrect, since
the complicated factor in the nummerator of $F_{qq}$
will not be canceled.

\section{Infinite order replica symmetry breaking}
Recently the author proposed a spherical model with infinite order replica
symmetry breaking. It is the model given by eq. (\ref{Ham=}), but summed
over $p$. \cite{Nqsg}
In particular, random pair ($p=2$) and random quartet interactions
($p=4$) occur, while random triplets should be absent. Then the
model has near the spin glass transition the same critical behavior
as the Sherrington-Kirkpatrick model. Its benefit is that it can be solved
explicitly in the whole low $T$ phase; this arises due to the spherical
nature of the spins and the mean field character of the model.
For our present purpose this allows an explicit check at any
temperature in a model with infinite RSB.

Assuming that the variance of the $p$-spin couplings equals
$J_p^2(p-1)!/N^{p-1}$, the free energy for a $k$-step RSB pattern
can be derived using results of an appendix of CS
\BEQ
\frac{\beta F}{N} =-\frac{1}{4}\beta^2 f(1)+\frac{1}{4}\beta^2\Phi
-\frac{1}{2}-\frac{q_0}{A_0}-\log(1-q_k)-\sum_{i=1}^{k}
\frac{1}{x_i}\log\frac{A_{i-1}}{A_i}
\EEQ
where
\BEQ f(q)=\sum_{p=2}^\infty \frac{J_p^2}{p}q^p\EEQ
and
\BEQ \Phi=\sum_{i=0}^k(x_{i+1}-x_i) f(q_i);\qquad
 A_i=1-q_k+\sum_{j=i+1}^kx_j(q_j-q_{j-1})\EEQ
As usual
we defined $x_0=0$ and $x_{k+1}=1$.
The variables $\{q_i,\, x_i\}$
will be denoted by a $2k+1$ component vector $v_i$, ($i=0,\cdots,2k$).
{}From $U=-(\beta N/2)(f(1)-\Phi)$ one obtains by differentiation
with respect to $T$ the specific heat
\BEQ\label{c2=}
 C=\frac{\beta^2N}{2}(f(1)-\Phi)+\frac{\beta^4N}{4}
\sum_{i,j=0}^{2k}\Phi_iR_{ij}\Phi_j\EEQ
where $R^{-1}_{ij}=\partial^2\beta F/\partial v_i\partial v_j$ and
$\Phi_i=\partial \Phi/\partial v_i$.

We again consider the fluctuations in the internal energy.
The analog of eq. (\ref{U1}) immediately explains the first two terms
of $T^2C$. The analog of eq. (\ref{19}) is
\BEQ \Delta U^{2\,(2)}=\frac{\beta^2N}{4}
\sum_{i,j=0}^{2k}\Phi_iR_{ij}\Phi_j\EEQ
Multiplied by $\beta^2$ this
 indeed equals the last term in eq. (\ref{c2=}).

Our approach thus reproduces the correct result for the fluctuations
of the internal energy at any fixed $k$ and, in particular, in the physical
limit $k\to \infty$.
In this continuum limit the meaning of the breakpoints $x_i$ becomes
ill-defined. It was conjectured by Parisi~\cite{GPpc} that the
discussed effect then disappears. In other words, the final result might
be due to fluctuations in the plateaus $q_i$ only.
In order to investigate this question we go to the
 Parisi free energy functional for the SK model~\cite{Parisi}
\BEQ \label{Parisimodel}
\beta F=\sum_{i=0}^k (x_{i+1}-x_i)
\left(\frac{\tau}{2}q_i^2-\frac{w}{3}q_iT_i+\frac{y_1}{8}q_i^4\right)
\EEQ
where the paramagnetic background has been subtracted,
$\tau=(\beta^2 J^2-1)/2$, and
\begin{equation}
T_i \equiv  \frac{1}{2}(2x_{i+1}-x_i)q_i^2+q_i\sum_{j=i+1}^k
(x_{j+1}-x_j)q_j+\frac{1}{2}\sum_{j=0}^{i-1}(x_{j+1}-x_j)q_j^2
\label{T_i}
\end{equation}
For this model eq. (\ref{c2=}) still applies with $f(q)=J^2q^2$,
while the term involving $f(1)=J^2$ comes from the paramagnetic background.

The nice thing of the Parisi model is that the solution
for $q_i$, $x_i$ is known exactly
at any $k$. It holds that
\BEQ q_i=\frac{2i+1}{2k+1}q_k\qquad x_i=i \frac{6y_1q_k}{w(2k+1)}\EEQ
where $q_k$ satisfies
\BEQ y_1q_k^2\left(\frac{3}{2}-\frac{1}{(2k+1)^2}\right)-wq_k+\tau=0\EEQ
This is an immediate, exact generalization of the Parisi
solution of the model (\ref{Parisimodel})
to the region where $\tau y_1/w^2$ is not very small.

{}From Parisi's results it follows that now
\BEQ U=\frac{\beta J^2N}{2}\sum_{i=0}^k (x_{i+1}-x_1)q_i^2
=\frac{\beta J^2N}{2}\left(q_k^2-\frac{8y_1k(k+1)}{w(2k+1)^2}q_k^3\right)
\EEQ
Differentiating with respect to $T$ one finds that
 $\Delta U^{2\,(2)}$ should be equal to
\BEA\label{duu} \Delta U^{2\,(2)}
=-\frac{\beta^2J^4Nq_k}{w}\,\,
\frac{w-y_1q_k(3-\frac{3}{(2k+1)^2})}
     {w-y_1q_k(3-\frac{2}{(2k+1)^2})}
\EEA
Since this expression converges as $1/(2k+1)^2$, a good
approximation is already obtained for $k=1$. The full result for the
specific heat converges even as $1/(2k+1)^4$.\cite{Parisi}

We have also calculated $\Delta U^{2\,(2)}$ by taking into
account only the fluctuations in the $q_i$. Numerically we
have inverted the matrix
$\partial^2 \beta F/\partial q_i\partial q_j $
up to large sizes ($k=150$) and find
at $w=y_1=1$, $\tau=0.1$ that
\BEQ
\sum _{i,j=0}^k
\frac{\partial \Phi}{\partial q_i}
\left( \frac{\partial^2 \beta F}{\partial q\partial q }\right)^{-1}_{ij}
\frac{\partial \Phi}{\partial q_j}
=\sum _{i,j=0}^\infty
\frac{\partial \Phi}{\partial v_i}
\left( \frac{\partial^2 \beta F}{\partial v\partial v}\right)^{-1}_{ij}
\frac{\partial \Phi}{\partial v_j}+ \frac{0.446}{(2k+1)^3}
=-0.4900 5929+ \frac{0.446}{(2k+1)^3}
 \EEQ
where the sum in the middle involves fluctuations
in both $q$ and $x$, viz. $\{v_i\}=\{q_0,q_1,q_2,\cdots; x_1,x_2,\cdots\}$.

{}From this example it is seen that Parisi's conjecture is probably true
in general:
the effect of fluctuations of the breakpoints disappears for continuous
order parameter functions.

\section{Conclusion}
We have put the basis for one-loop calculations
in a mean field spin glass model with 1RSB.
We calculated the energy fluctuations because their strength is known
already; they are proportional to the the specific heat.
In order to reproduce the correct answer,
 we had to take into account fluctuations in both the
height and the location of the breakpoint of the 1RSB solution.
(It should be admitted that the normalization
of the fluctuation integral deserves a  solid derivation.
Our assumed form is the natural one, and leads to the correct results.)
\cite{Ferrero}

It is the new point of this paper that in integrating over all
$q_{\alpha\beta}$ one must not only integrate over the
fluctuations around the saddlepoint at given parametrization
(given breakpoint $x_1$), but also
over all possible parametrizations (locations of the breakpoint).
In hinsight,
this is not so unnatural. In mean field both the location
of the breakpoint and the plateau value are determined by a
saddle point equation. Generally saddle points appear when one
has to perform sharply peaked integrals. The very notion
of a saddle point can only occur due to smallness of fluctuations
around it. Saddle points without fluctuations do not exist.
Fluctuations must always be considered; only in some cases they
are negligible.

The new question which arises is: what happens beyond mean field
in finite range systems? Physically, the weight of the 1RSB solution
is the probability for occurrence of non-zero overlap. At before hand,
there is no reason why it should be constant. One would thus
expect that it fluctuates.  This is exactly what we showed in an
explicit mean field calculation. Beyond mean field one expects
that it fluctuates in space. In the renormalization group approach of
Cwilich and Kirkpatrick~\cite{Cwilich}
such fluctuations are not taken into account.

An important question is whether they are compatible with ultrametricity.
The answer is affirmative. Indeed, given Parisi's division of the replica
number $n=(n/x_1)(x_1/x_2)\cdots(x_k/1)$, one can impose at
each lattice sites the allowed ``ultrametric states'' $x_1,\cdots,x_k$.
In mean field only one ``state'', say $\bar x$, will be occupied;
this happens uniformly in space. For describing fluctuations
of the breakpoint at site $r$,
the state with breakpoint $x(r)=\bar x+\delta x(r)$ will be occupied
at the cost of the occupation of the state with $\bar x$.
This will modify the longitudinal sector of the fluctuations.
As this sector is massive at the transition and below,
it is expected to bring only quantitative effects, and not to change
the renormalization flow.~\cite{Cwilich}.

In models with infinite order replica symmetry breaking
the order parameter function has in zero field a continuous
part up to $x=x_1$ and is constant beyond. Here the effect
of fluctuations in the location of the breakpoints was seen
to disappear. For short range systems this implies that the
breakpoints donot fluctuate in space. The fluctuations of
 the probability of the self-overlap,
i.e. $1-x_1$, are probably already taken into
account by the fluctuations in the height of the plateau.

\acknowledgments
The author benefited from participation in the
conference `` Low temperature dynamics and phase-space structure
of complex systems'', held at Nordita, Kopenhagen, March 1995.
He also acknowledges discussion with J.M. Luck and
hospitality at the CEA Saclay, where part of the work was done.
He thanks Giorgio Parisi for urging him to publish this material.
This work was made possible by the Royal Dutch Academy of Arts
and Sciences (KNAW).

\end{document}